\documentclass [12pt]{article}
\usepackage{amsmath}
\usepackage{epsfig}
\usepackage{amsfonts}
\begin{document}
\def\b{\bar}
\def\d{\partial}
\def\D{\Delta}
\def\cD{{\cal D}}
\def\cK{{\cal K}}
\def\f{\varphi}
\def\g{\gamma}
\def\G{\Gamma}
\def\l{\lambda}
\def\L{\Lambda}
\def\M{{\Cal M}}
\def\m{\mu}
\def\n{\nu}
\def\p{\psi}
\def\q{\b q}
\def\r{\rho}
\def\t{\tau}
\def\x{\phi}
\def\X{\~\xi}
\def\~{\widetilde}
\def\h{\eta}
\def\bZ{\bar Z}
\def\cY{\bar Y}
\def\bY3{\bar Y_{,3}}
\def\Y3{Y_{,3}}
\def\z{\zeta}
\def\Z{{\b\zeta}}
\def\Y{{\bar Y}}
\def\cZ{{\bar Z}}
\def\`{\dot}
\def\be{\begin{equation}}
\def\ee{\end{equation}}
\def\bea{\begin{eqnarray}}
\def\eea{\end{eqnarray}}
\def\half{\frac{1}{2}}
\def\fn{\footnote}
\def\bh{black hole \ }
\def\cL{{\cal L}}
\def\cH{{\cal H}}
\def\cF{{\cal F}}
\def\cP{{\cal P}}
\def\cM{{\cal M}}
\def\ik{ik}
\def\mn{{\mu\nu}}
\def\a{\alpha}
\title{Gravitating lepton bag model}

\author{Alexander Burinskii, \\
 Theor. Phys. Lab., NSI, Russian Academy of Sciences,\\
 B. Tulskaya 52,  Moscow 115191 Russia, e-mail: burinskii@mail.ru}

\maketitle

\begin{abstract}
\noindent As is known, the gravitational and electromagnetic (EM) field of the Dirac electron is described by an over-extremal  Kerr-Newman (KN) black hole (BH)  solution which has the naked singular ring and two-sheeted topology. This space is regulated by  the formation of a regular source based on the Higgs mechanism of broken symmetry. This source shares much in common with the known MIT- and SLAC-bag models, but has the important advantage, of being in accordance with gravitational and electromagnetic field of the external KN solution. The KN bag model is flexible. At rotations, it takes the shape of a thin disk,  and similar to other bag models, under deformations it creates a string-like structure which is positioned along the sharp border of the disk.
\end{abstract}

{11.27.+d, 04.20.Jb, 04.60.-m, 04.70.Bw}

\maketitle

\section{Introduction and Overview}
It has been discussed for long time that black holes (BH) are to
be related with elementary particles \cite{BHpart}. The Kerr-Newman (KN) rotating BH solution paid in
this respect especial interest, since,
 as it was shown by Carter \cite{Car}, its gyromagnetic ratio $g=2$ corresponds to the Dirac electron,
 and therefore, the four measurable parameters of the electron:
spin $J$, mass $m$, charge $e$ and magnetic moment $\mu$ indicate that
gravitational and electromagnetic field of the electron should be described by the KN solution.
   In the recent paper \cite{DokEr}  Dokuchaev and Eroshenko considered a solution of the Dirac
   equation under  BH horizon, and
    suggested that this model may represent a ``...particle-like charged solutions in general relativity...''. On
   the other hand, it should be noted that the model of  a Dirac particle confined under BH horizon can also be
   considered as a type of gravitating bag model,  and it acquires especial interest since this bag is to
be gravitating, leading to a progress beyond the known MIT and SLAC bag models \cite{MIT,SLAC}.
However, the spin
and charge of elementary particles are very high with respect to their masses, which prevents formation of the BH horizons. In particular, the Kerr-Newman solution with parameters of  electron: charge $e$, mass $m ,$ and spin parameter $a=J/m$ exceeds the threshold value $e^2 +a^2 \le m^2$ for existence of the horizons about 21 order.  Similar ratios for the other elementary particles show that besides the Higgs
boson, which has neither spin nor charge, none of the elementary particles may be associated with a true black hole, and rather, they should be associated with the over-rotating Kerr geometry, with $|a|>>m .$

The corresponding over-rotating KN space has topological defect -- the naked Kerr
singular ring, which forms a branch line of space
 into two sheets described by different metrics: the sheet of advanced and sheet of the retarded fields. The Kerr singular and related two-sheeted structure created the problem
of a mysterious source of the Kerr and KN solutions, which paid
considerable attention  during more than four decades
\cite{Keres,Isr,Ham,Lop,Bur0,IvBur,GG,BurSol,BurSol1}. For the
story of this investigation, see for example \cite{BEHM}. Long-term
attempts to resolve the puzzle of the source of Kerr geometry led
first to the model of the vacuum bubble -- the rotating disk-like
shell \cite{Lop,Ham}.
In the subsequent works, the vacuum state inside the bubble turned into a superconducting bulk formed of a
condensate of the  Higgs field, \cite{BurSol,BurSol1}. The structure of source acquired typical
features of the soliton and Q-ball models, becoming similar to the known bag models \cite{MIT,SLAC}.

 Recent analysis of the
Dirac equation inside the KN soliton source \cite{BurDirKN}, confirmed that the regularized KN solution shares much
in common with the known MIT and SLAC bag models. However, the \emph{gravitating bag} formed by the KN bubble-source
should have specific features  associated with the \emph{need to preserve the external KN field}.

On the other hand, the semiclassical theory of the bag models \cite{SLAC} includes elements of quantum
theory which are \emph{based on a flat space-time} without gravity, and we are faced with a known conflict between gravity
and quantum theory. Our solution to this problem in \cite{BurSol,BurSol1}  is based on two requirements:
\begin{description}
         \item[I:] The space-time should be flat inside the bag,
         \item[II:] The space-time outside the bag should be the exact KN solution.
       \end{description}
Therefore the quantum-gravity conflict is resolved by separation of their regions of influence. Remarkably,\emph{ these requirements determine features of the KN bag unambiguously.
}
First of all, they  determine uniquely border of the KN bag, showing explicitly that, in accord with general concept of the bag models \cite{SLAC,Giles}, the KN bag has to be flexible and its shape depends on the rotation parameter $a=J/m ,$ as well as from the local intensity of the electromagnetic (EM) field.

As a result, for parameters of an electron, the rotating  bag takes the shape of a thin disk of ellipsoidal form, see Fig.1.
 Its thickness $R$ turns out to be equal to classical radius of electron $ r_e =e^2/2m ,$ while radius of the disk corresponds to the Compton wave-length of the disk,\fn{This was determined by L\`opez \cite{Lop}.} which allows to identify it with a \emph{dressed electron}.

 The the degree of oblateness of this disk is  $a/R = \alpha^{-1} = 137 ,$   the fine structure constant $\alpha$ acquires a geometric interpretation.

\begin{figure}[ht]
\centerline{\epsfig{figure=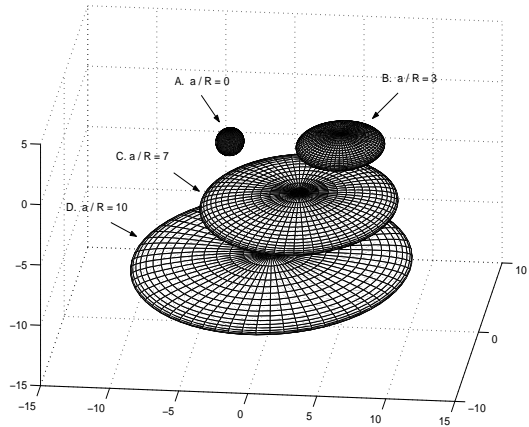,height=6cm,width=7cm}}
\caption{(A): Spherical bad with zero rotation, $a/R =0$, and the rotating disk-like bags for different ratios $a/R$,  (B): $a/R =3$;(C): $a/R =7$, (D):$a/R =10$.}
\label{fig3}
\end{figure}

 The next very important consequence of these requirements is the emergence of a ring-string structure on the bag border,
 \cite{Bur3Q}, and further the emergence of a singular pole associated with traveling-wave excitations of the string \cite{BurQ}. This pole can be associated with a single quark, and finally, the KN bag takes the form of a coherent ``bag-string-quark'' system.

Finally, these requirements  determine that the Higgs condensate should be enclosed \emph{inside the bag,} contrary to the standard treatments of the bag as  \emph{a cavity in the Higgs condensate}, \cite{MIT}.
 Realization of this requirement cannot be done with the usual quartic potential for self-interaction of the Higgs fiield
 \cite{MIT,SLAC}, and requires a more complicate field model, based on a few chiral fields and a supersymmetric scheme of phase transition \cite{WesBag}.

At this point we have to mention the important role of Kerr Theorem, which determines the null
vector field $ k_\m (x) ,$  the Kerr Principal congruence which forms a vortex polarization of
Kerr-Schild (KS) metric \be g_\mn =\eta_\mn + 2H k_\m k_\n . \ee The Kerr theorem gives two solutions
for this congruence $k_\m^\pm ,$ which determine two sheets of the KN solution corresponding to two
different metrics $g_\mn^\pm .$ Solutions of the Dirac equation on the KN background should be
consistent with the metric corresponding to one of this congruence.

We show that two solutions of the
Kerr theorem generate two massless Weyl spinor fields which are coupled into a Dirac field
consistent with the Kerr geometry. However, the null spinor fields of the Kerr congruences are
massless, and there appears the question on the origin of the mass term. Answer comes from theory
of the bag models \cite{SLAC}), where the Dirac mass is a variable depending on the local vev of
the Higgs condensate.

This gives a direct hint to a consistent embedding of the  Dirac equation in
the regularized KN background, indicating that the both sheets of the KN solution are necessary as
\emph{carriers of the initially massless leptons}. It turns out in agreement with the basic
concepts of the Glashow-Weinberg-Salam model \cite{GSW}, in which the lepton masses are generated
by the Higgs mechanism of symmetry breaking.

As a result, we conclude that two-sheeted Kerr's structure is an essential element for the
consistent with gravity space-time realization of the electroweak sector of the Standard Model.

\section{Over-rotating Kerr geometry: two-sheeted structure and regular source}
The KN solution in the Kerr-Schild (KS)  form \cite{DKS} has the metric \be g_\mn =\eta_\mn + 2H
k_\m k_\n ,\label{KSH} \ee where $ \eta_\mn $ is metric of
auxiliary Minkowski space, $x^\m =(t.x.y.z) \in M^4 ,$\fn{We use signature $(- + + +)$.} and
\be H= \frac {mr-e^2/2}{r^2 + a^2 \cos^2 \theta} \label{H}.\ee

 The vector field $ k_\m $ is null, $ k_\m k^\m =0 ,$ and determined by the differential form
\be k =k_\m dx^\m = dr - dt -a \sin^2 \theta d\phi ,\label{k} \ee
where $t, r, \theta, \phi ,$ are the Kerr oblate spheroidal coordinates
\be x +iy =(r +ia) e^{i\phi} \sin \theta, \quad z= r\cos \theta , \quad t =\rho - r . \label{obl coord} \ee

The field $ k^\m (x) $ forms Principal Null Congruence
(PNC) \cite{BoyLink}, $\cal K ,$ which determines polarization of the Kerr space-time, see Fig.2.
The PNC is focussed at the Kerr singular ring,  $r=0, \ \cos \theta=0 ,$ which is the branch
line of the Kerr space into two sheets $r>0$ and $r<0 .$\fn{These are Riemannian sheets of the Kerr complex
radial distance $\tilde r = r+ ia \cos \theta .$}

Extension of the Kerr congruence to negative
sheet of the KS space ($r<0$) along the lines $\phi =const., \
\theta=const.$  creates  another congruence with different
radial direction, and the congruence which is \emph{outgoing} by $r>0$ turns on the negative sheet into
\emph{ingoing} one.\fn{The relations (\ref{obl coord}) are
changed too, \cite{BoyLink}.} Thus, the
Kerr solution describes in the KS form two different sheets of
space-time, determined by two different congruences \be k_\m^{\pm}(x)dx^\m = \pm
dr - dt -a \sin^2 \theta d\phi \label{kpm} \ee and two
different metrics \be g_\mn^\pm =\eta_\mn + 2H k_\m^\pm k_\n^\pm
\label{KSpm} \ee on the same Minkowski background $x^\m \in M^4.$

This two-sheetedness  created the problem
of the source of Kerr geometry, and there appeared two lines of investigation.
One of them, \cite{Bur0, IvBur,BurSen}, accepted two-sheetedness as
indication of its plausible connection with a spinor structure of the
Kerr space-time and with two-sheeted structure of the topologically
nontrivial ``Alice'' strings introduced by Schwarz and Witten \cite{Wit}.

Alternative line of the investigation was related with truncation of the
KN negative sheet, and with a consistent replacement of the excised region
by a source in agreement with the Einstein-Maxwell field equations, \cite{Keres,Isr,Ham,Lop,GG,BurSol,BurSol1}.
\begin{figure}[ht]
\centerline{\epsfig{figure=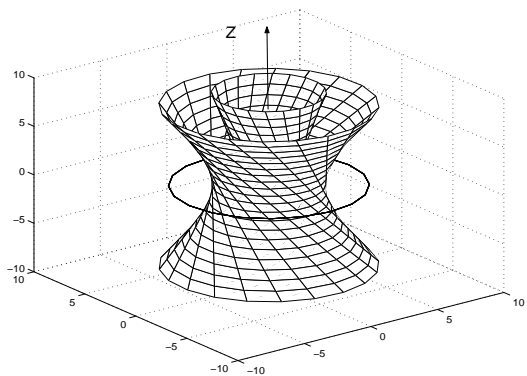,height=5cm,width=7cm}}
\caption{Kerr's principal congruence of the null lines (twistors)
is focused on the Kerr singular ring, forming a branch line of the
Kerr space into two sheets. }
\end{figure}

There is a freedom in chouse of the truncating surface, and in the most successful version
of the model suggested by L\'opez \cite{Lop}, the KN source formed a bubble, boundary
of which was determined by matching the external KN metric (\ref{KSH}) with a flat metric inside the bubble.
According to (\ref{KSH}) and (\ref{H}),  this boundary has to be placed at the radius
 $ r = R = \frac {e^2}{2m} .$

From (\ref{obl coord}), one sees that $r$ is indeed the oblate spheroidal coordinate determined by the equation
\be \frac {x^2+y^2}{r^2 + a^2} + \frac {z^2}{r^2} =1 ,\ee
 and the source of KN solution takes the form of a very oblate disk of radius
 $ r_c \approx a = \frac {1}{2m}$ with thickness
 \be r_e = e^2/2m \label{re} ,\ee
which is the classical radius of electron. So, that the fine structure constant acquires geometrical meaning as the degree of oblateness of the disk-like source, $r_e/r_c = e^2 \approx 137^{-1}.$

As a result of the regularization, the disk-like region surrounding the Kerr
singular ring is excised and replaced by flat space,
which acts as a cut-off parameter -- an effective minimal distance $R=r_e $ to the former Kerr singular ring.
Note that for the case without rotation, $a=0 ,$ the disk-like bubble takes the
spherical form and size of classical  electron (\ref{re}).

The L\'opez model was later on transformed into a \emph{soliton-bubble} model
\cite{BurSol,BurSol1}, in which the thin shell of the bubble was replaced by a field model of a
domain wall providing a smooth phase transition between the external KN solution and the flat
internal space. This phase transition was modelled by Higgs mechanism of broken symmetry, and flat
interior of the KN bubble was formed by a supersymmetric state of the Higgs condensate.

 The field model of broken symmetry is similar to Landau-Ginzburg
 model of superconductivity \cite{NO}, and regularization of the
 singular KN solution can be viewed as an analogue to the Meissner
 effect, expulsion of the gravitational and EM fields from interior of the
 superconducting source.

\section{Higgs condensate and mass of the Dirac field}

The used for regularization of the KN solution Higgs mechanism of broken symmetry relates
the source of KN solution with many other extended particle-like models of the electroweak sector
of the standard model. In particular, with the
 superconducting string model of Nielsen and Olesen \cite{NO,AchVach}, with  Coleman's Q-ball
 models, \cite{GRos,Coleman,Kusen,VolkWohn,Grah} and with the famous MIT- and SLAC- bag models.
 In this paper we pay especial attention to  fermionic sector of the KN source and obtain close
 similarity between the Higgs mechanism of mass generation  in the KN soliton model and that
 in the SLAC bag model \cite{SLAC}.

Hamiltonian of the SLAC model for interaction of the  Higgs field with the Dirac field $\psi$ has
the form \be {\cal H} =
\int d^3 x \{  \psi^\dag(-i \vec{\alpha} \cdot \vec\nabla +  g \beta \sigma)\psi +
\frac 12 (\dot \sigma ^2 + |\vec\nabla \sigma|^2) + V(\sigma) \} ,\ee
where $g$ is a dimensionless coupling parameter, and self-interaction of the non-linear Higgs field
$\Phi$ is described by quartic potential
\be V(|\Phi|)=g(\bar\sigma \sigma - f^2)^2 ,\label{Vf4} \ee
where $\sigma =<|\Phi|>$ is  vacuum expectation value (vev) of the Higgs field.
The true vacuum state of the Higgs field $\sigma =0$ is not the state of lowest energy, and
the Higgs field is triggered in the state  $ \sigma = f ,$  which breaks
the gauge symmetry of the spinor field $\psi .$ As a result, fermion acquires the mass $m=g\eta $
which is used in the confinement mechanism of the bag models.
However, the condensate of the Higgs field $ \sigma = f $ breaks also the gauge symmetry of
the EM fields. In the known bag models, it turns
the external EM fields in the short-range one which distorts the external KN solution.

For example, in the MIT-bag model the Higgs vev vanishes inside the bag, $r<R , $ and takes nonvanishing
value $ \sigma = f ,$ in
 \emph{ outer region,} $r> R .$ See. Fig.3.
\begin{figure}[ht]
\centerline{\epsfig{figure=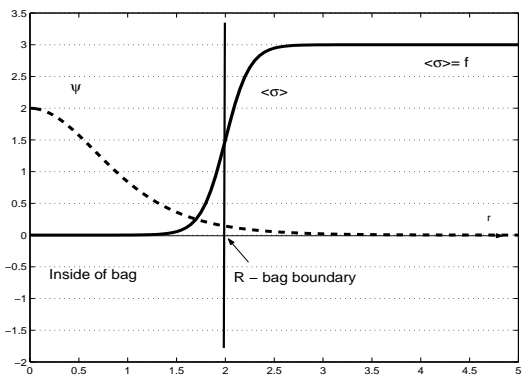,height=5cm,width=7cm}}
\caption{Positions of the vev of Higgs field $\sigma$ and the confined spinor wave function $\Psi$
(quark) in the MIT-bag model.}
\end{figure}

The Dirac equation  in the presence of the $\sigma$-field takes the form \be (i\gamma ^\m \d_\m -
g\sigma) \psi =0 \label{Dir-sigma} , \ee
and the Dirac wave function $\psi $  turns out to be massless inside the bag and  acquires a large
effective mass $m=g f $ outside.   The quarks are confined inside the bag, where they have the most
energetically favorable  position.

Geometry of the Higgs vacuum state is different in the SLAC bag models, see Fig.4.
The vev $\sigma$ gives the  mass to  Dirac field outside the bag as well as inside.
The mass vanishes only in the very narrow region near the surface of the bag, $ r \sim R .$

\begin{figure}[ht]
\centerline{\epsfig{figure=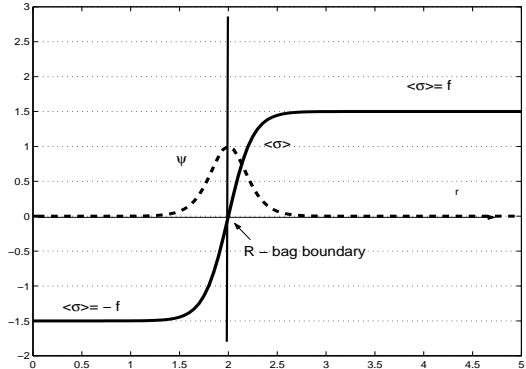,height=5cm,width=7cm}}
\caption{Classical solutions of the SLAC-bag model. The vacuum
field $\sigma $ and the localized spinor (quark) wave function
confined to the thin shell - boundary of the bag.}
\end{figure}

Such geometry of the broken vacuum state creates a  sharp localization of the Dirac wave function at the
 border  of the bag.

 In the bag models we are faced  with several very important novelties:

  (A) The statement on impossibility of localization of the Dirac wave function beyond the distances
  comparable with the Compton wave length $\hbar / mc $  is violated, and quarks can localize
  within a very thin region at the bag shell. The reason of that is scalar nature of the confinement potential,
  for which ``...there is no Klein paradox of the familiar type encountered in the presence of strong,
  sharp vector potential.'' \cite{SLAC}.

(B) There is effectively used a semi-classical approach to  one-particle Dirac theory.  Solving Dirac
equation for a quark in scalar potential assumes that all the negative-energy states are filled, and
treatment is focused on the lowest positive-energy eigenvalues. Therefore, ``...there is  no ambiguity
in identifying and interpreting the desired positive energy "one-particle" solutions.''\cite{SLAC,DrellExp-Th}

(C) Mass term of the Dirac equation (\ref{Dir-sigma}) is determined by the vev of Higgs field
$\sigma(x) = <|\Phi(x)|>$, and therefore, it  turns out to be a  function in configuration space.

(D) Bag models presumed to be very soft,  compliable and extensible. They are easily deformed, and   under
rotations and deformations they may acquire extended stringy structures accompanied by vibrations \cite{Giles}.

All these peculiarities of the bag models are compatible with the soliton-bubble source of KN solution.
However,
there is one important difference: the typical bag model represents a bubble or cavity in a superconducting
media -- Higgs condensate, while  in the gravitating bubble-source of the KN solution the Higgs condensate
is enclosed \emph{within the bubble}, leaving unbroken the true vacuum outside the bag.

 In the MIT and SLAC bag models, the Higgs condensate is placed outside the source, and the external vacuum
 represents a superconducting state, which leads to the short-range external
 EM fields.

  A dual (turned inside out) geometry was suggested in the Coleman  Q-ball model \cite{Coleman}.  The self-interacting Higgs of a Q-ball is confined inside the ball-like source, $ r<R ,$ leaving the external vacuum unbroken.  Most of the Q-ball models led to a coherent oscillating state of the Higgs vacuum inside the bag\fn{Such a model was first considered by G. Rosen, \cite{GRos}.}(oscillons \cite{Kusen,VolkWohn,Grah}). The KN soliton-source \cite{BurSol,BurSol1} exhibits also this
  peculiarity.
  We can summarize that \emph{confinement of the Higgs condensate inside the bag is necessary  requirement  for  the correct gravitating properties of the bag models}.
However, formation of the corresponding potential turns out to be a very non-trivial problem, which cannot be solved by the usual quartic potential (\ref{Vf4}).

\section{Field model of broken symmetry and phase transition for gravitating bag model}

Among theories with spontaneous symmetry breaking, important place takes the field model of  a vortice in condensed matter which was considered by Abrikosov in connection with theory of type II superconductors. Nielsen and Olesen (NO) used this solution for a semiclassical relativistic string model \cite{NO}. The NO string model, representing a magnetic flux tube in superconductor, was generalized to many other semiclassical field models of the solitonic strings and found wide application in the electroweak sector of the standard Glashow-Salam-Weinberg (GSW) model \cite{AchVach,SchifYan}.

The NO model \cite{NO} contains a complex scalar field $\Phi$ and the gauge EM field $A^\m$ which becomes massive through the Higgs mechanism. The Lagrangian has the form
 \be
{\cal L}_{NO}= -\frac 14 F_\mn F^\mn - \frac 12 (\cD_\m
\Phi)(\cD^\m \Phi)^* - V(|\Phi|), \label{LNO}\ee where $ \cD_\m =
\nabla_\m +ie A_\m $  are the $U(1)$ covariant derivatives, and $F_\mn
= A_{\m,\n} - A_{\n,\m} $ is the field strength.
The potential $V$ has the same quartic form as in (\ref{Vf4})
\be V = \lambda ( \Phi ^\dag \Phi - f^2)^2 ,\ee
where $\sigma$ is replaced by complex field $\Phi =|\Phi| e^{i\chi} .$

 The Lagrangian ${\cal L}_{NO}\equiv {\cal L}_{mat}$ describes a vortex string
 embedded in the superconducting Higgs condensate \emph{in flat space-time}.
 Similarly to the bag models, this model cannot be generalized to gravity, since the
 Higgs condensate gives mass to the external  EM  field, turning it into a
 non-physical short-range field conflicting with the external KN solution.

Improvement of this flaw was suggested by  Witten  in his $U(1)\times \tilde U(1) $ field
model of a cosmic
superconducting string \cite{Wit}, in which he used two Higgs-like fields, $\Phi^1 $ and $\Phi^2 $.
One of them, say $\Phi^1 ,$ had the required behavior, being concentrated inside the source, while
another one, $\Phi^2 ,$ played auxiliary role and took the external complementary domain extending up
to infinity. These two Higgs field are charged and adjoined to two
different gauge fields $A^1$ and $A^2 ,$ so that when one of them
is long-distant in some region $\Omega$, while the second one is
long-distant in complimentary region $\overline {\Omega}= U_\infty \setminus \Omega.$ This model is
suitable for any localized gravitating source, however, for superconducting source of the KN
solution we used in
\cite{BurSol} a supersymmetric generalization of the Witten model suggested by Morris \cite{Mor}.

\subsection{Supersymmetric phase transition}
The supersymmetric scheme of phase transition is based on
three chiral fields $\Phi^{(i)}, \ i=1,2,3 ,$  \cite{WesBag}. One of this fields, say $\Phi^{(1)} ,$
has the required radial dependence, and we chose it as the Higgs field ${\cal H} ,$
setting the additional notations in accord with
$({\cal H}, Z, \Sigma) \equiv (\Phi^0, \Phi^1, \Phi^2) .$

The coupled with gravity action should reads
\be S = \int \sqrt{-g} d^4 x(\frac R {16 \pi G} + {\cal
L}^{mat}) ,\ee
where the full matter Lagrangian takes the form
 \be {\cal L}^{mat}= -\frac 14 F_\mn F^\mn - \frac 12
\sum_i(\cD^{(i)}_\m \Phi^{(i)})(\cD^{(i) \m} \Phi^{(i)})^* - V
\label{L3} ,\ee which contains contribution from triplet of the
chiral field $\Phi^{(i)}.$

 The required for our model potential $ V$  is obtained  by a
 standard supersymmetric scheme of broken symmetry \cite{WesBag}, which determines it via superpotential
 $W(\Phi^i, \bar \Phi^i)$,
 \be V(r)=\sum _i |\d_i W|^2  .\ee
 The superpotential leading to the required geometry of broken symmetry was suggested by J.Morris
 \cite{Mor}: \be  W(\Phi^i, \bar \Phi^i) = Z(\Sigma \bar \Sigma -\eta^2)
+ (Z+ \m) {\cal H} \bar {\cal H},\label{W} \ee where $ \m, \
\eta $ are real constants. It yields
\be V=(Z+\m)^2 |{\cal H}|^2 + (Z)^2 |\Sigma|^2 +
(\Sigma \bar \Sigma + {\cal H} \bar {\cal H}-\eta^2  )^2 , \label{VPhi}\ee

\noindent and equation \be \d_i W =0 \ee determines two vacuum
states separated by a spike of the potential $V$ at $r \approx R$:

\begin{figure}[ht]
\centerline{\epsfig{figure=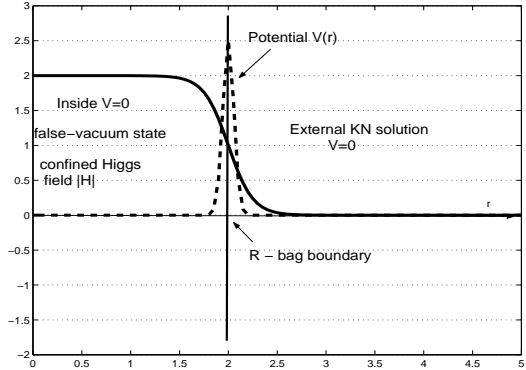,height=5cm,width=7cm}}
\caption{Region of broken symmetry in the KN soliton bag model.
The potential $V(R)$ forms the inner and outer vacuum states $V=0$
with a narrow spike at the boundary of the bag. The Higgs field $H$ is
confined inside the bag, $r<R$, forming a supersymmetric condensate which gives a mass
 to the Dirac equation.}
\end{figure}

\medskip

\textbf{EXT:} external vacuum, $r>R +\delta $, $V (r) = 0 ,$ with vanishing
Higgs field ${\cal H} =0 $,  while $Z =0, \Sigma =\eta ,$ and

 \medskip

\textbf{INT:} internal vacuum state, $r<R -\delta $, $V (r) = 0$ , while $|{\cal H}| = \eta , \ \ Z=-\m , \ \Sigma = 0  $).
\fn{This state could be called a false-vacuum, since vev of the Higgs field is non-zero. However, the term false-vacuum has already been used in the literature in a different sense, and we use here the term "Higgs condensate". Howeve, the term false-vacuum was used in our previous paper \cite{BurDirKN}.}

 \subsection{Application to KN source}

 The choose of  L\'opes's boundary for regularization of the KN source allows us to neglect gravity inside the source and at the boundary, and thus, \emph{one can neglect gravitational field in the zone of phase transition and consider the space-time as flat}.  In the same time, outside the source we have exact Einstein-Maxwell gravity, since the gauge symmetry is unbroken and all the terms $ \frac 12 (\cD_\m
\Phi)(\cD^\m \Phi)^*$ vanish together with the potential $ V(|\Phi|).$ Therefore, outside the source we have only the matter term ${\cal L}^{mat}) \equiv -\frac 14 F_\mn F^\mn $ leading to the external KN solution.

 Therefore, inside the source (zone \textbf{INT}) and on the boundary, we have only the part of Lagrangian which corresponds to self-interaction of the complex Higgs
field and its interaction with vector-potential of the KN electromagnetic field $A^\m $ in the flat space-time.

 The field model is reduced to the model, considered by Nielsen-Olesen for a vortex string in the superconducting media \cite{NO}, \be
{\cal L}_{NO}= -\frac 14 F_\mn F^\mn - \frac 12 ({\cal D}_\m {\cal
H})({\cal D}^\m {\cal H})^* - V(|{\cal H}|), \label{LNO}\ee where
$ {\cal D}_\m = \nabla_\m +ie A_\m $ is a covariant derivative,
$F_\mn = A_{\n,\m} - A_{\m,\n}  ,$  and $\nabla_\m \equiv \d_\m$ is reduced to derivative
in flat space with a flat D'Alembertian $\d_\n \d^\n = \Box  .$
For interaction of the complex Higgs field \be{\cal H}(x) = |{\cal H} (x)| e^{i\chi(x)} \ee
with the Maxwell field we obtain the
following complicated systems of the nonlinear differential equations
\bea D_\n D^\n {\cal H} &=& \d_{\bar{\cal H}} V  , \label{PhiIn} \\
\Box A_\m = I_\m &=&  e |{\cal H}|^2
(\chi,_\m + e A_\m). \label{Main} \eea

 \medskip

\noindent
The obtained vacuum states \textbf{EXT} and \textbf{INT} show that $ |{\cal H}(r)|$ should be a step-like function
\be |{\cal H}(r)| =\begin{cases}
\eta & \text{if $r \le R-\delta $},\\
0& \text{if $r \ge R+\delta$}.
\end{cases} \label{Hr}
 \ee
with a transition region $ R -\delta < r < R+\delta $ where its behavior is determined by the impact of electromagnetic field.

Outside the source, $r>R+\delta$, we have ${\cal H} =0 $ and obtain $I_\m = 0 .$
Inside the source, by $r\le R-\delta ,$ we have also $ I_\m = 0 ,$  which is provided there by compensation of the vector potential by a gradient of the phase $\chi$ of the Higgs field, $\chi,_\m +e A_\m =0 .$ So, the nonzero current exists only in the narrow transitional region $ R -\delta < r < R ,$ where this compensation is only partial, and (\ref{Main}) describes a ``region of penetration'' of the EM field inside the Higgs condensate, see Fig.5.

\subsection{Important consequences}

 Analysis of the equation (\ref{Main}) in \cite{BurSol,BurSol1} showed two
 remarkable properties of the KN rotating soliton:

 \medskip

 \textbf{(I)} The vortex of the KN vector potential $A_\m$ forms a quantum
 Wilson loop placed along the border of the disk-like source, which leads to
 \emph{quantization of the angular momentum} of the soliton,

\textbf{(II)} the Higgs condensate should \emph{oscillate} inside the source
with the frequency $\omega= 2m $.

The KN vector potential has the form \cite{DKS}
\be A_\m dx^\m = - Re \
[(\frac e {r+ia \cos \theta}) (dr - dt - a \sin ^2 \theta d\phi
) \label{Am}. \ee
Maximum of the potential is reached in the equatorial plane,  $\cos \theta =0 $,
at the  L\'opez's boundary of the disk-like source (\ref{re}), $r_e = e^2/2m ,$
which plays the role of a cut-off parameter,
\be A^{max}_\m dx^\m = - \frac e {r_e} (dr - dt - a d\phi) \label{Amax}. \ee

 The $\phi-$ component of vector potential, $A^{max}_\phi = ea/r_e ,$ shows that the potential
forms near the source boundary a circular flow  (Wilson loop). According to (\ref{Main}),
this flow is compensated inside the soliton by  gradient of the Higgs phase $\chi,_\phi ,$
and does not penetrate inside the source beyond a transition region $r< r_e -\delta .$
Integration of this relation over the closed loop $\phi =[0,2\pi]$ under condition
$I_\phi =0$ yields the result \textbf{(I)}.

Similarly, using (\ref{Main}) and condition $I_\phi =0$
for the time component of the vector potential $A^{max}_0  = \frac e {2 r_e} = m/e ,$
we obtain the result \textbf{(II)}.

\section{Fermionic sector of the KN bag model}
Now we have to consider matching of the solutions of Dirac equation with interior of the
regular solitonic source and with external KN solution. We start from the region inside
the KN source and the adjoined $\delta$-narrow layer of phase transition, $r < R+\delta .$
 In accord with the used scheme of regularization, these regions are to be flat, and one
 can use here the usual Dirac equation $\gamma^\m \d_\m \Psi = m \Psi ,$ which
in the Weyl representation splits into two equations
\be
 \sigma ^\m _{\alpha \dot \alpha} i \d_\m
 \bar\chi ^{\dot \alpha}=  m \phi _\alpha , \quad
 \bar\sigma ^{\m \dot\alpha \alpha} i \d_\m
 \phi _{\alpha} =  m \bar\chi ^{\dot \alpha},
\label{Dir} \ee
where the Dirac bispinor $\Psi = \left(\begin{array}{c}
 \phi _\alpha \\
\bar\chi ^{\dot \alpha}
\end{array} \right)$ is presented by two Weyl spinors $ \phi _\alpha $ and $
\bar\chi ^{\dot \alpha}.$

In the conception of the bag models, fermions acquire mass via a Yukawa coupling to the Higgs
field,  (\ref{Dir-sigma}), and since the Higgs condensate in the KN source is concentrated
inside the bag, (\ref{Hr}), the mass term of the Dirac equation  takes the maximal value
\be m=g\eta  \label{mbare}\ee in internal region, while outside the bag the Dirac equation turns out to
be massless, and splits into two independent massless equations
\be
 \sigma ^\m _{\alpha \dot \alpha} i \d_\m
 \bar\chi ^{\dot \alpha}=0 , \quad
 \bar\sigma ^{\m \dot\alpha \alpha} i \d_\m
 \phi _{\alpha} = 0,
\label{Dir0} \ee
corresponding to the left-handed and right-handed ``electron-type leptons'' of the
Glashow-Salam-Weinberg model \cite{GSW}.

Outside the bag we have an external  gravitational and EM
fields of the KN solution, and one should use the Dirac equation in covariant form
\be  \gamma^\m_{KS} {\cal D}_\m \Psi =0, \label{DirKS} \ee
where $\gamma^\m_{KS} $ are $\gamma$-matrixes adapted to the Kerr-Schild form of metric
(\ref{KSH}), and \be{\cal D}_\m =\d_\m - \frac 12 \Gamma_{\n\lambda \m} \Sigma^{\n\lambda}
-i \frac k {2\sqrt 2} \gamma_\m F_{\n\lambda}\Sigma^{\n\lambda} \label{calD}\ee are covariant derivatives.

The exact solutions on KS background were earlier considered by S. Einstein and R. Finkelstein
in \cite{EinFinkel}, and following them we can choose the $\gamma^\m_{KS} $ matrixes in the form
\be \gamma^\m_{KS} = \gamma^\m_{W} + \sqrt {2H} k^\m \gamma^5_{W} \label{gamKS} ,\ee
where $\gamma^\m_{W}$ are matrixes of the Weyl representation for Minkowski space $\eta^\mn .$
They satisfy the usual anticommuting relations
\be \{ \gamma^\m_{W}, \gamma^\n_{W} \} = 2 \eta^\mn,
\quad \{ \gamma^\m_{W}, \gamma^5_{W} \} =0, \quad (\gamma^5_{W})^2 =-1 ,\ee
while  $\gamma^\m_{KS}$ satisfy the anticommuting relations
\be \frac 12 \{ \gamma^\m_{KS}, \gamma^\n_{KS} \} = \eta^\mn - 2H k^\m k^\n = g^\mn_{KS} , \ee
adapted to KS metric.
It is known that the exact KS solutions belong to the class of the algebraically special solutions,
for which all the tensor quantities are to be aligned with Kerr null congruence, \cite{DKS}, and
the general relations (\ref{DirKS}), (\ref{gamKS}), (\ref{calD}) become
much simpler when the Dirac field $\Psi(x)$ is ``aligned'' with the Kerr  congruence $k^\m(x)  ,$
\be k_\m \gamma^\m \Psi =0 . \label{alignDir}\ee
 For the aligned Dirac field, the nonlinear terms of the electromagnetic and gravitational
 interactions are cancelled, and the Dirac equation linearized \cite{EinFinkel},
 taking the form of a free Dirac equation in flat space-time (\ref{Dir0}).

The alignment condition (\ref{alignDir}) may be rewritten in the form
\be
(\vec k \cdot \vec \sigma)
 \phi = \phi, \quad
(\vec k \cdot \vec \sigma)
 \bar\chi = - \bar\chi   , \quad
 \label{Dir0pm} \ee
which shows that the left-handed and the right-handed fields
$\bar\chi $ and $\phi $ are to be oppositely polarized with respect to space direction of
the Kerr congruence $\vec k .$
We obtain that only one of these two ``half-leptons'', the left-handed $\phi$ is really
consistent with the Kerr congruence $k^+=(1,\vec k),$ selected for the \emph{physical sheet} of
the KN solution. The consistent solution takes the form $\Psi_L^T =(\phi,0),$ which shows
explicitly that only left-handed field $\phi$ is aligned with $k^+$ and survives on the physical
sheet of the KN geometry. This solution is exact, since for the massless Dirac equation
the left- and right-handed spinors are independent. Similarly, we obtain the solution
$\Psi_R^T =(0,\bar\chi),$ which is not aligned with $k^+$ and with the selected physical
sheet of the KN solution. However, it is aligned with congruence $k^-$ and can ``lives'' on the
negative sheet of advanced fields.
Thus, the massive Dirac solution $\Psi = \left(\begin{array}{c}
 \phi _\alpha \\
\bar\chi ^{\dot \alpha}
\end{array} \right)$ splits into the left and right massless parts $\Psi_L$ and $\Psi_R$,
which \emph{outside the bag} can live only on the different sheets of the twosheeted Kerr
geometry.

This important peculiarity of the Dirac solutions on the Kerr background was also mentioned
in \cite{EinFinkel}, where authors noted  that the Dirac equations on the KS background
``\emph{...are not consistent unless the mass vanishes...}''.
Meanwhile, this obstacle disappears inside the bag-like source of the Kerr geometry, where the
space is flat by construction of the solitonic source (sec.2). When the massless
Weyl spinors pass from two different external sheets on a common flat space inside the bag,
they are combined into a Dirac bispinor which gets mass from the Higgs
condensate through Yukawa coupling (see Fig.6).
Removing the two-sheeted structure, which was associated with the problem the source of KN solution,
we meet its appearance from another side, by analysis of the consistent solutions of the Dirac
equation on the KS background. We obtain that two-sheeted structure of the KS geometry is agreed
with elementary constituents  of the standard model -- the massless`left-handed' and `right-handed'
electron fields \cite{GSW,DrellExp-Th}, providing consistency of the external Dirac field with KN gravity.

The Kerr congruences are determined  by \emph{the Kerr theorem},
\cite{DKS,BurTMP}, which is formulated in twistor term on the auxiliary to KS metric
(\ref{KSH})  Minkowski space $\eta_\mn .$ The first twistor component, $Y$ plays also the role of
a projective spinor coordinate (see details in Appendix and \cite{BurDirKN,BurTMP}).
The Kerr theorem gives for the KN particle two solutions $Y^\pm (x)$ which are connected by
antipodal relation $Y^+ =- 1/\bar Y^-$ and determine two antipodal congruences
$k^+_\mn(x) $ and $k^-_\mn(x) .$  The Weyl spinors corresponding to solutions $Y^\pm(x)$ are
exactly the considered above Weyl spinor components $\phi$ and $\bar\psi$ of the aligned Dirac solutions.
Since the Kerr theorem is formulated in the flat space-time, the solutions $Y^\pm (x)$ are
extended unambiguously from the external KN space to the flat space inside the bag, which determines
 the Dirac bispinor
\be\tilde \Psi = \left(\begin{array}{c}
 f_1 (x)\phi _\alpha \\
f_2 (x)\bar\chi ^{\dot \alpha}
\end{array} \right), \label{fPsi}\ee which is aligned  to the both external congruences and represents a
constraint, selecting in the flat space inside the bag the Dirac solution with required polarization.

\begin{figure}[ht]
\centerline{\epsfig{figure=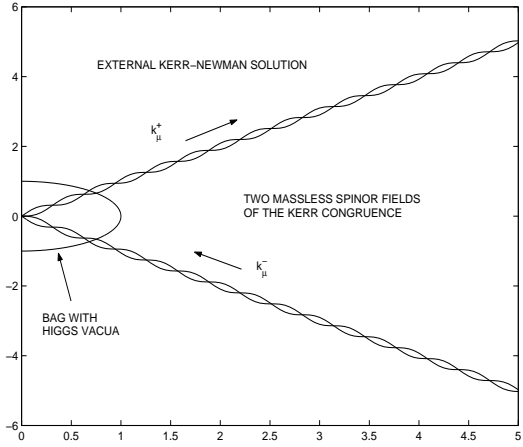,height=6cm,width=7cm}}
\caption{Two sheets of external KN solution are matched with flat space inside the bag. The
massless spinor fields $\phi_\alpha$ and $\bar\chi^{\dot\alpha} $ live on different KN sheets,
aligned with $k^+_\m$ and $k^-_\m $ null directions.  Inside the bag they
 join into a Dirac bispinor, which obtains the mass from the Higgs condensate concluded inside the bag.}
\end{figure}

Another very specific peculiarity of the  bag models is  emergence of
the variable mass term in the Dirac equation (\ref{Dir-sigma}).  The mass therm is
determined by the vev of
 Higgs condensate $\sigma$ which depends on the regions of space-time, and
 in the region of the maximum of the Higgs condensate $\sigma=\eta,$ is called the
the bare mass $m =g\eta.$
The Dirac wave function, solution of the Dirac equation with variable mass term, avoids
the region with a large bare mass, and tends to get a more energetically
favorable position, which is principal idea of the quark confinement.

In  the SLAC bag model \cite{SLAC}, the resulting wave function is determined by variational
approach. The  Hamiltonian is
\be H(x) =\Psi^\dag (\frac 1{i} \vec \alpha \cdot \vec \nabla
+g\beta\sigma ) \Psi \label{Ham}, \ee and the energetically
favorable wave function is determined by minimization of
the averaged Hamiltonian ${\cal H} = \int d^3 x H(x) $
under  the normalization condition $\int d^3x \Psi^\dagger(x) \Psi(x) =1 .$
It yields
\be (\frac 1{i} \vec \alpha \cdot \vec \nabla
+g\beta\sigma ) \Psi = {\cal E} \Psi \label{H-E}, \ee
where ${\cal E}$ appears
as the Lagrangian multiplier enforcing the
normalization condition.
Similar to results of the SLAC-bag model, one expects that the Dirac
wave function will not penetrate deeply in the region of large bare mass
$m =g\eta ,$ and will concentrate in a very narrow transition zone at the
bag border $R-\delta <r< R+\delta $.  As it was motivated in \cite{SLAC},
the narrow concentration of the Dirac wave function is admissible in the bag
models, since for the scalar potential there is no the Klein paradox.
 The exact solutions of this kind are known only for
two-dimensional case, and the corresponding variational problem
 should apparently be solved numerically by
using the ansatz (\ref{fPsi}), where
 $f_1(x)$  and $f_2(x)$ are
variable factors.

 The use of classical solutions of the Dirac equation in a given scalar potential leads also
 to the problem of the negative-energy states.  In the bag models  this problem is considered
 semiclassically by using the assumption \cite{SLAC} that ``...all the negative-energy states
 in the presence of this potential are filled...'', and as a result, it is necessary to
 consider only the lowest positive-energy eigenvalues.\fn{This is an approximation to rigorous treatment
 based on the normal-ordering. The negative-energy states
 correspond to the charge-conjugate solutions
 $ \Psi^c (x) = C\bar\Psi(x), \quad \bar\Psi^c (x)= C^{-1} \bar\Psi(x) $ of the charge-conjugate
 Dirac equations (with replacement $e \to -e ,$). In particular, the density of current is determined by
 the commutation relations
 $ j_\m (x) = \frac  {ie} 2 [\bar\Psi(x), \gamma_\m \Psi(x)] =\frac  {ie} 2 (\bar\Psi(x)
 \gamma_\m \Psi(x) - \frac  {ie} 2 \bar\Psi^c (x) \gamma_\m \Psi^c(x)) ,$ and similar for the
expectation value of energy or any other operator bilinear in fermion field.}

Splitting of the Kerr-Schild space-time \emph{outside the source} of KN
solution looks strange from the point of view of the standard gravitation, but it appears
more natural by comparison with electromagnetism, which is sensitive to difference between
the retarded and advanced fields.

It is known, \cite{MTW}, that the Kerr solution may be represented in the Kerr-Schild (KS)
form via the both Kerr congruences $k^+_\m $ or $k^-_\m ,$ but not via the both simultaneously.
For the KN solution with EM field, situation is more complicated. Although the both representations are
admissible, the representation via retarded fields is physically preferable, since the
asymptotic advanced EM field of the KN solution will contradict to its experimental behavior
 in flat space.
 Vector potential $A_\m$ of the KN solution must also be aligned with the Kerr congruence, and
 should be retarded, $A_{ret},$ on the physical sheet determined by the outgoing Kerr congruence
 $k^+_\m .$ The appearance of advanced EM fields, $A_{adv} $ is important in the non-stationary problems.
 In particular, in the Dirac theory of radiation reaction, the retarded
  potentials $A_{ret}$ are split into a half-sum and half-difference
with advanced ones $ A_{ret} = \frac 12
[A_{ret} + A_{adv}] + \frac 12 [A_{ret} -
A_{adv}],$ where \be A_{ret}^+ = \frac 12 [A_{ret} + A_{adv}] \ee is connected
  with radiation reaction, and  \be A_{ret}^- = \frac 12 [A_{ret} - A_{adv}] \ee forms a self-interaction of the source. Similar
structure presents also in the Feynman propagator.

  The fields $A_{ret}$ and $A_{adv}$ cannot reside on the same physical sheet of the Kerr geometry, because each of
  them should be aligned with the corresponding Kerr congruence. Considering the retarded sheet
  as a basic physical sheet, we fix the congruence $k^+_\m $ and the corresponding metric  $g_\mn^+$, which are not allowed for the advanced field $A_{adv}$ and must be positioned on the separate sheet which different metric  $g_\mn^- .$

\section{Discussion.}

Taking the bag model conception, we should also accept their dynamical properties that they
soft and easily deformed \cite{SLAC,Giles}, forming a stringy structure. Typically, these are radial and rotational excitations
accompanied by the formation of the open tube-shaped string ended with quarks. Another type of deformations was considered in the  Dirac model of an "extensible" electron (1962) \cite{DirBag}, which can also be regarded as a prototype bag
model with radial excitations.\fn{This view  was also suggested in \cite{Tarak}.
Interpretation of the black holes and AdS geometries as a sort of the bag was also noticed in
\cite{Derig,Bever}.}  The bag-like source of the KN solution without rotation, $a = 0 ,$ coincides
with this "extensible" model of the Dirac electron leading to  the ``classical electron
radius'' $R=r_e=e^2/2m .$ As we discussed in Introduction, the disk-like bag of the rotating  KN source can be viewed as stretching
of the spherical bag by rotations. For parameters of an electron, the spinning bag stretched
in a disk of radius, $a= \hbar/2mc ,$ covering the Compton area of ``dressed'' electron.
 The disk is very thin with degree of flattening $ \alpha = 137 ^{- 1} .$ The boundary of disk
appears to be very close to former position of Kerr singular ring, and the  EM field near the
boundary may be seen as a regularization of the KN singular EM field. Similar to other singular lines, the
Kerr singular ring was considered as a string in \cite{IvBur}. The structure of the EM field
 near this string was analyzed first in \cite{Bur0,IvBur}, and much later in \cite{BurSen}.
 It appeared to be similar to the structure of the fundamental string solution, obtained by Sen to
 low-energy heterotic string theory \cite{BurSen}. It is a  typical light-like pp-wave string
 solution \cite{HorSteif,BurQ}, which in the Kerr geometry takes the ring-like form.

 Regularization of the KN source does not remove this ring-string, but gives
it a cut-off parameter (\ref{re}), $R=r_e$. It was shown in \cite{Bur0,IvBur} and later specified in
\cite{BurQ,BurA,BurTMP} that the EM excitations of the KN solution lead to appearance of  traveling
waves propagating along this ring-string. However, the light-like ring-string cannot be closed \cite{BurDStr},
since the points different by angular period,   $x^\mu (\phi, t)$ and  $x^\mu (\phi+2\pi, t)$
should not coincide, and a peculiar point on the ring-string should make it open, forming a single
quark-like  endpoint.

The string traveling waves deform the bag boundary creating singular pole \cite{Tomsk}.
We will not discuss  it here in details, leaving the treatment to a separate paper.
Note only, that the exact solutions for the EM excitations on the Kerr background were obtained in  \cite{DKS},
and using the considered in introduction conditions \textbf{I}, and \textbf{II} we can  unambiguously determine
the back reaction of the local EM field on the metric, and obtain the corresponding deformations of the bag boundary.
The origin of singular pole is caused by a circulating node in the EM string excitation. This node yields the zero
cut off parameter $R$, creating contact of the bag boundary with singular ring.

This singular pole circulates along the sharp border of the disk with
the speed of the light and may be considered three-fold: a) as a light-like quark enclosed
inside the bag, b) as a single end-point of the light-like ring-string (as it was shoved in
\cite{BurDStr}, the light-like fundamental string cannot be closed), and c) as a naked point-like
electron enclosed in a circular `zitterbewegung'. It leads to an integrated model for the dressed
and bare electron as a single coherent system similar to the hadronic bag models.
\section{Conclusion}
Starting from  the old problem of the source of the KN geometry we obtained first a
bubble-core model of the spinning particle,  the supersymmetric vacuum state of which is formed by the Higgs mechanism of symmetry breaking.
Contrary to the most other known models of the particle-like objects, the KN bubble forms a \emph{gravitating} soliton  creating the external gravitational and EM field of an electron. This compatibility with gravity has required the use of a supersymmetric field model of phase transition leading to a supersymmetric vacuum state in the core of the particle,  leaving unbroken the external EM field.

 The resulting soliton model has much in common with the famous MIT and SLAC bag models,  but gets the ``dual bag geometry'', in which the Higgs condensate is embedded ``inside out ''  compared to  previous bag models.

The  two-sheeted structure of the Kerr geometry has got in this model a natural space-time
 (coordinate) implementation forming a background for the initially massless leptons of the Glashow-Salam-Weinberg model \cite{GSW}.

Without pretension on the detailed description, we can note that the described dressed electron may
be turned in a positron, if we  change the role of the advanced and retarded sheets of the Kerr
geometry. The higher excitation of the ring-string may generate the muon state, while switching off
the scalar and longitudinal components of the EM field, corresponding to charge of the KN solution
\cite{TNieu}, and preserving only the transversal traveling waves, gives us a neutral particle,
which has the features of a neutrino. Therefore, some variations the KN bag model can give us the
space-time structure for some other spinning particles of the electroweak sector of the Standard
Model.

 \textbf{Acknowledgments.} This
work is supported by the RFBR grant 13-01-00602. Author thanks
 Theo M. Nieuwenhuizen, Yuri Rybakov and Bernard
Whiting for interest and useful conversations in the early stages of  this work.
Author would like to thank P. Kondratenko and Yu. Obukhov and all colleagues of the Theor. Phys. Laboratory of NSI RAS for useful discussion, and also V. Dokuchaev, V. Rubakov and other members of  Theoretical Division of INR RAS for invitation to seminar talk and useful discussion. Author thanks J. Morris for reading this paper, very useful conversation and pointing out some mistakes in signs.  The results of this work were also recently delivered at the International Conference ``Solitons: Topology, Geometry, and Applications'' in Thessaloniki, Greece, and author would like to thank organizers for invitation and all  participants, in particular T. Manton and B. Schr\"oers, for useful conversations. This research is supported by the RFBR grant 13-01-00602.

\section*{Appendix. The Kerr theorem} The Kerr theorem determines all the geodesic and \emph{shear free}
 congruences as analytical solutions of the equation
\be F(T^A) =0 ,\label{FTA}\ee where $F$ is an arbitrary
holomorphic function of the projective twistor variables \be T^A=
\{ Y, \ \z - Y v, \ u + Y \Z \}, \qquad A=1,2,3 , \label{(TA)} \ee
where $\z = (x+iy)/\sqrt 2, \ \Z = (x-iy)/\sqrt 2 , \ u = (z +
t)/\sqrt 2 , \ v = (z - t)/\sqrt 2 $ are  null Cartesian
coordinates of the auxiliary Minkowski space.

We notice, that the first twistor coordinate $Y$ is also a
projective spinor coordinate \be Y =\phi_1/\phi_0 , \label{Y10}
\ee and it is equivalent to two-component Weyl spinor $\phi_\alpha
,$ which defines the null direction\fn{We use the spinor notations
of the book \cite{WesBag}, where the $\sigma$-matrixes has the form
$\sigma^\m =(1, \sigma^i), \ \bar\sigma^\m =(1,  - \sigma^i), \ i=1,2,3 $ and
$\sigma^\m =\sigma^\m_{\alpha \dot \alpha} ,\ \bar\sigma^\m
=\bar\sigma^{\m \dot\alpha \alpha} $.}
   $ k_\m = \bar
\phi_{\dot\alpha} \sigma_\m^{\dot\alpha \alpha}\phi_\alpha .$

It is known, \cite{DKS,BurTMP}, that function $F$ for the
Kerr and KN solutions may be represented in the quadratic in $Y$
form, \be F(Y,x^\m) = A(x^\m) Y^2 + B(x^\m) Y + C(x^\m).
\label{FKN} \ee In this case  (\ref{FTA}) can explicitly be
solved, leading to two solutions
 \be Y^\pm (x^\m)= (- B \mp \tilde r )/2A, \label{Ypm}\ee
where $\tilde r= (B^2 - 4AC)^{1/2} .$ It has been shown in \cite{BurTMP}, that these solutions are antipodally
conjugate, \be Y^+ = - 1 /{\bar Y^-} \label{antipY}.
\ee

Therefore, the solutions (\ref{Ypm}) determine two Weyl spinor
fields $\phi_\alpha $  and $\bar\chi_{\dot\alpha}$, which in
agreement with (\ref{antipY}) are related with  two antipodal
congruences \be Y^+ = \phi_{1}/\phi_{0} \label{Y+10} ,\ee \be Y^-
= \bar\chi_{\dot 1}/\bar\chi_{\dot 0} \label{Y-10} .\ee
 In the Debney-Kerr-Schild (DKS) formalism \cite{DKS} function $Y$ is
also a \emph{projective angular coordinate} $Y^+ = e^{i\phi} \tan
\frac \theta 2 .$ It gives to spinor fields $\phi_\alpha$ and
$\bar\chi_{\dot\alpha}$ an
 explicit  dependence on the Kerr angular coordinates $\phi$ and $\theta .$

For the congruence $Y^+ $ this dependence takes the form \be \phi_{\alpha} =
\left(\begin{array}{c}
e^{i\phi/2} \sin \frac \theta 2   \\
e^{-i\phi/2} \cos \frac \theta 2
\end{array} \right) . \label{phiY+}\ee
In agreement with (\ref{antipY}) we have $\Y^- = - e^{-i\phi} \cot
\frac \theta 2 ,$ and from invariant normalization
$\phi_\alpha \chi^\alpha = 1$ we obtain $ \chi_{\alpha} =
\left(\begin{array}{c}
 - e^{i\phi/2} \cos \frac \theta 2   \\
 e^{-i\phi/2} \sin \frac \theta 2
\end{array} \right)$ which yields \be \bar\chi^{\dot\alpha}
= \epsilon ^{\dot\alpha \dot\beta} \bar\chi_{\dot\beta} =
\left(\begin{array}{c}
 e^{i\phi/2} \sin \frac \theta 2   \\
 e^{-i\phi/2} \cos \frac \theta 2
\end{array} \right) \label{phiY-} .\ee

\end{document}